# Antenna System for Simultaneous Wireless Power and Information Transfer to Brain Implants

Ali Khaleghi, Aminolah Hassanvand, Ilangko Balasingham
Department of electronic systems (IES), Norwegian University of Science and Technology, Trondheim, Norway
*ali.khaleghi@ntnu.no*

***Abstract*— Brain-Computer Interfaces (BCIs) have revolutionized neuroscience applications, from motor rehabilitation to neuroergonomics. Traditional implantable BCIs with invasive microelectrode arrays pose challenges, notably the need for wired connections and inherent implantation risks. This paper introduces a battery-free wireless BCI system, consolidating an implant and its external supporting system. Our design centers on a dual-function antenna system: firstly, an inductive coupling mechanism enables wireless power transfer, sufficiently powering the implant's Application-Specific Integrated Circuit (ASIC) for stimulation and readout without an implant battery. Secondly, a backscatter antenna in the implant facilitates battery-free, high-data-rate wireless connectivity (up to 32 Mbps). This system not only enhances the BCI experience by eliminating wires but also retains data fidelity and energy efficiency, promising a safer, more efficient interface for tasks like robotic arm control.**

***Index Terms*— Antenna Design, Brain-Computer Interfaces (BCIs), Inductive Coupling, Wireless Communication, Wireless Power Transfer.**

## I. INTRODUCTION

Brain-Machine Interfaces (BMIs) have revolutionized the field of neural engineering by facilitating direct communication between the brain and external devices. This communication is achieved through various techniques, including non-invasive methods like electroencephalography (EEG), implantable surface electrocorticography (ECoG), and micro-electrode arrays. Among these, implantable electrode arrays stand out, having transformed neural research by enabling direct interfacing with neural tissue and allowing high-resolution neural recording and stimulation. Prominent examples of implantable electrode arrays include the Utah Electrode Array (UEA) by Blackrock Neurotech [1], Neuronexus Probe by NeuroNexus, Silicon Probe Arrays by Atlas Neuroengineering [2], Polytrode Arrays by Cambridge NeuroTech, and NeuroPixels. These devices have showcased their prowess in high-density neural recording. They have proven particularly adept at decoding neural activity related to motor intentions and sensory perception. With multi-site recording capabilities, these arrays permit simultaneous monitoring of neural signals from different brain areas. Such comprehensive recording facilitates the development of advanced neuroprosthetic devices, expanding applications from neurorehabilitation to motor function restoration and robotic control. Systems like the NeuroPort by Blackrock Microsystems now offer detailed recordings across numerous channels at high frequencies. These systems capture intricate neural activity, enriching our understanding of neural processes and aiding the creation of accurate decoding algorithms. However, these probes typically require wired connections through a headstage on the skull, with signal amplification and data sampling executed outside the body.

The integration of wireless technology could enhance implantable electrode array capabilities, but power constraints for high-speed data remain a challenge [3]. Our proposition is to employ wireless backscatter for data transmission, thereby eliminating the implant's active transceiver system. By supplementing this with wireless power through magnetic coupling, we aim to establish a battery-free, high-speed implantable BMI system. Such a system could mitigate risks tied to wired connections and pave the way for prolonged recording, closed-loop systems, and advanced BMIs.

This paper outlines our developments under the H2020 FET-OPEN B-CRATOS European Project. We seek to establish seamless brain connectivity to external devices in a feedback loop for precise control of a prosthetic hand and sensory cortex stimulation in a feedback loop. This paper details the antenna subsystem's integration with the implant unit and support system, ensuring robust data transmission for both readout and telemetry functions of the implant.

## II. SYSTEM DESCRIPTION

The overall concept and configuration of the BMI implementation, which is the focus of this work, is illustrated in Fig. 1. A Utah microelectrode array is set to be implanted in the motor cortex and somatosensory areas for non-human primate (NHP) experiments. The array is connected to the implant's integrated ASIC, which is responsible for neural signal amplification, channeling, and analog-to-digital (ADC) conversion. The 32-channel Utah electrode array, with a sampling rate of 30 KHz per channel, delivers an accumulated data rate of 32 Mbps for neural recording and parallel stimulating channels. High-rate sampling ensures more accurate intention recording, facilitating the precise control of external devices. The generated data controls the implant's integrated backscatter antenna through a binary RF switching mechanism. This switching alters the implant antenna's radar cross-section (RCS). Consequently, an external reader, operating at a radio frequency of 434 MHz, can detect changes in the RCS. This process extracts the ASIC-produced data from the motor cortex, providing the motor information in the external reader. By utilizing this approach, the implant's power consumption for data transmission is reduced to the energy required by the RF switch, which consumes only 165 nW. The main power consumption within the implant comes from the ASIC, estimated to be in the range of 20-30 mW. Power can be supplied wirelessly through an external source using induction coupling, resulting in minimal power dissipation in the tissues. The wireless powering coil, integrated with the backscatter antenna, makes the implant more compact. Overall, the implant unit operates without a battery, relying on external power. It can simultaneously transmit motor data to the reader and receive stimulation data via the magnetic link.

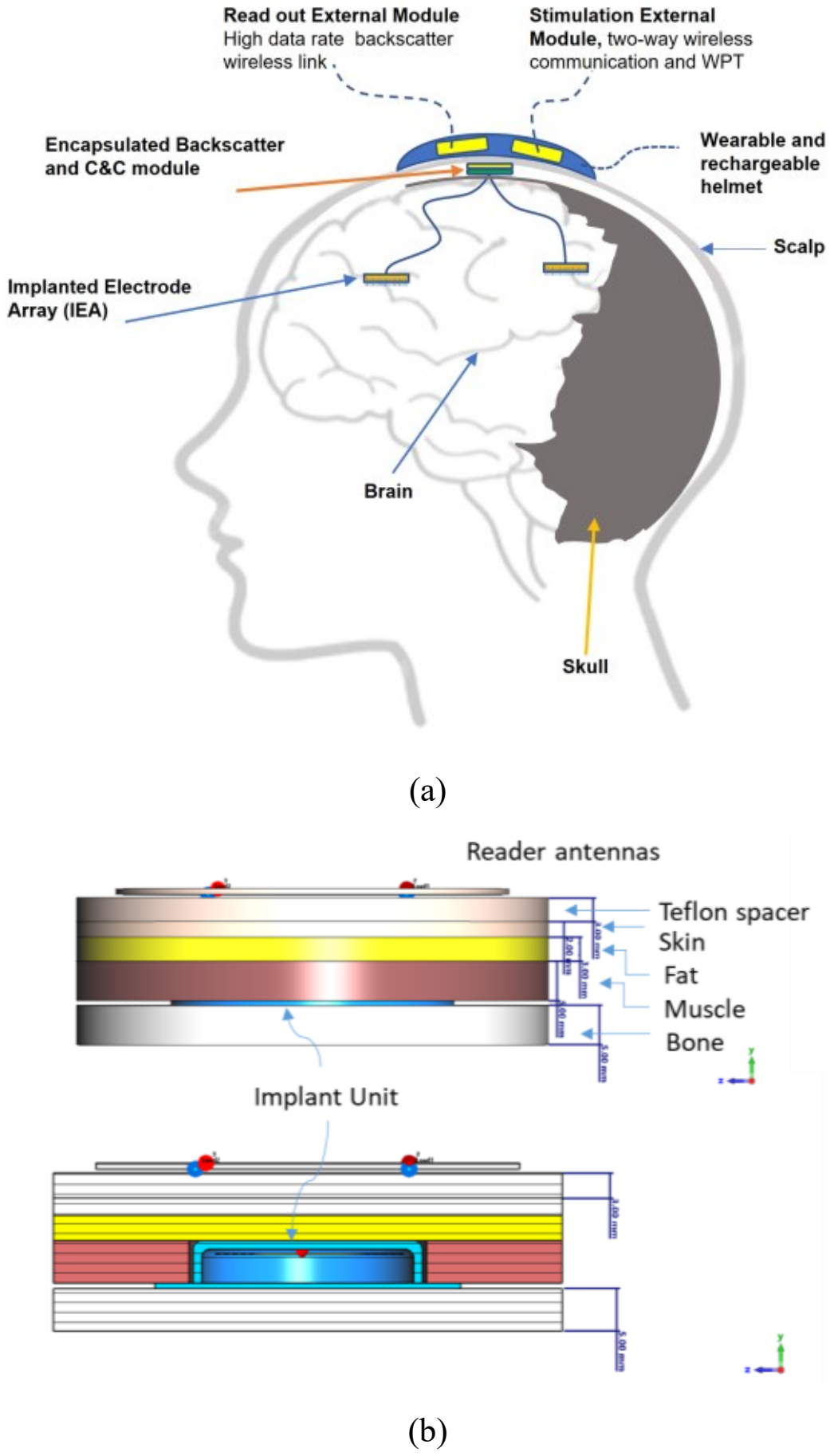

Fig. 1. a) Brain Machine Interface (BMI) with a wearable wireless power and wireless backscatter units b) Simulation model of the implant and external units in a layered tissue

In our design, the antenna system comprises two stacked implant antennas. The first is a coil designed to receive magnetic flux, connected to a wireless management unit that regulates programmed voltage and current directed to the ASIC. The second is a cross dipole, enhanced with multiple patches and equipped with a switch for RCS alternation. The external unit features a coil intended for wireless power transmission to the implant. Additionally, it includes two bi-static reader antennas with minimal mutual coupling, optimized for RF backscatter at 434 MHz.

### *A. Induction coil design*

We have developed a magnetic induction system to power the device. To fit within the ceramic enclosure of the implant, which is a cylinder with a diameter of 25mm and a thickness of 7mm, we have selected planar coils. As depicted in Fig.1-b the modeled electromagnetic simulation system incorporates layered tissues: skin and fat covering the top of the implant, and muscle that mimics the fluids surrounding the implant, terminating at the base where the implant adheres to the surface of the skull. The distance of the implant surface to the head surface is only 5 mm as the enclosure is placed under the skin. Electromagnetic modeling uses the finite element method (FEM) within the frequency domain at 13.56 MHz. The biological tissues are modeled using single frequency material properties [4]. This frequency aligns with the standard for the near-field coupling (NFC) protocol. Figure.2a shows the 3D model of the coil and the implant where the integrated implant coil is also illustrated To enhance precision, the simulation is also conducted using a heterogeneous model of the human head, ensuring accurate representation of both the layers and inherent curvatures (Fig. 2 b).

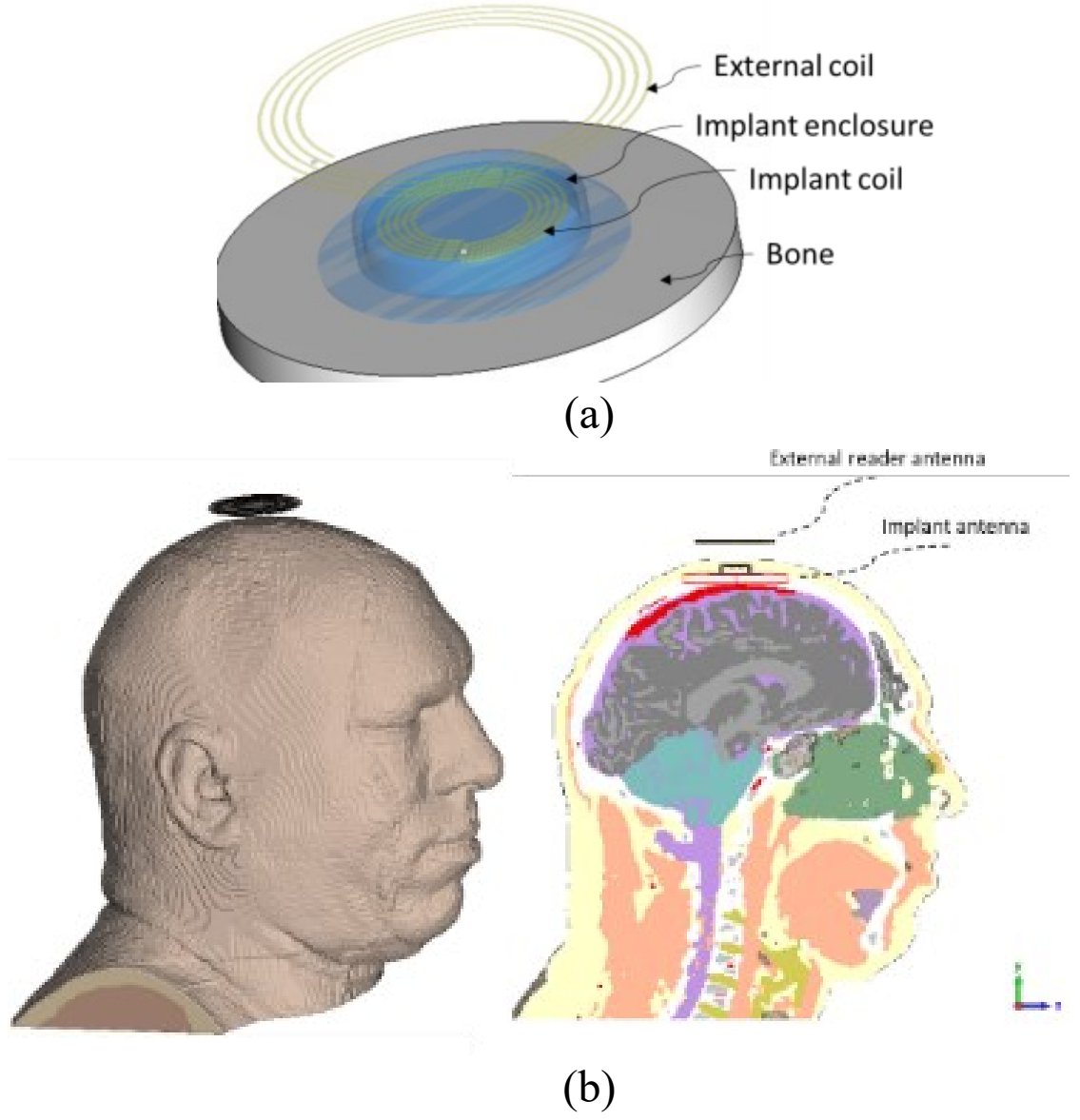

Fig. 2. a) Positioning of the external antenna and implant coil c) Simulation model using a realistic human representation.

At 13.56 MHz, the impedance measurements are as follows: the implant impedance 0.77+j48, and the external unit impedance is 1.47+j100. With an inter-coil distance of 14 mm, the ideal power coupling is obtained to be -7 dB with a matching Q-factor of 14. To counteract the effects of frequency splitting—resulting from the close proximity of the coils—and maintain power transfer efficiency, we've incorporated some loss into the matching circuit, leading to a revised Q-factor of approximately 5. The calculated SAR for magnetic induction assuming 500mW accepted power is SAR (10g) 0.12 W/Kg for multilayer tissues and 0.2 W/kg for the heterogenous model. In practice, the applied power is reduced according to the implant power consumption with maximum 300 mW using the telemetry link from the implant, thus lesser SAR is expected. We note that the standard SAR (10g) <2 W/kg [5] for general public users is recommended.

In practical implementation the implant antenna is matched to the NTAG 5 NXP powering unit where single capacitor is used to match the antenna. The data rate for this communication and powering unit is maximum 500 Kbps, where the bandwidth is sufficient to handle the data connectivity.

In the context of system integration with the backscatter communication (as discussed in section B), there is notable interference from NFC in the RF backscatter link due to intermodulation effects. To mitigate this, a displacement of 2 cm is introduced between the wireless power transfer (WPT) and the backscatter reader antennas. Although this adjustment limits the power handling capacity to the implant, but it adequately supports the up to 30 mW of power needed by the ASIC chip. As the separation distance increases, the resonance split effect becomes less pronounced, allowing for a higher Q-factor matching. It's crucial to arrange a balance in the system configuration to minimize the impact of both systems.

## *B. Implant and External Antennas*

The backscatter reader employs a bi-static radar approach, transmitting an unmodulated continuous wave (CW) RF signal and receiving reflections from the proximate implant antenna. Specifically, the transmitter system emits a CW RF in the ISM band at 434 MHz. A second antenna receives a portion of this emission as the coupling and the reflected data due to RCS changes. Given the simplicity of the implant electronics, which lacks coding and error correction in its data format, it is vital to minimize the mutual coupling between the transceiver antennas to detect the weak data signal. When antennas are situated closely, the coupling between them can be significant, necessitating innovative approaches. A prime solution is to employ different polarizations and spacing between the transmitter and receiver antennas. The antenna design shown in Fig. 3 uses two meander loop antennas with near field intensity in the meander area. The coupling between the antennas is -17 dB after matching at 435 MHz. The meander line extends the antenna physical length in the available small area and provides the sufficient coupling in the near field to sense the backscatter implant alternations.

The coupling between the external antennas and the implant at 15 mm is approximately -27 dB (using the model in Fig. 2) and independent of the implant rotation around its vertical axis. The antenna is simulated using heterogenous model of human head model. The mutual coupling of the reader antennas is slightly reduced to about -22 dB. The antenna SAR (10g) is calculated for the RF system using the multilayer tissue model, and using the heterogenous human head model at 434 MHz as shown in Fig.4. The maximum SAR (10g) is below 2.7 W/kg using the multilayer tissue on the skin, and it is below 1.66 W/kg using the heterogeneous model for accepted power of 1 W. The difference is related to the geometry and the shape difference in the models and the effective distance to the head surface. We note that the applied power for communication is 100 mW that reduces the continuous SAR to 0.27 W/kg and 0.166 W/kg for the multilayer and heterogeneous model. The small SAR permits safe use of the RF backscatter for BMI continuous application. The thermal effects in the electronics should be evaluated that would be the main thermal limiting factor.

The aforementioned antenna system is used in lab prototypes to demonstrate the feasibility of simultaneous wireless power and high data rate communication using inductive coupling and Rf backscatter. The results of this part are presented in [6] where the data rate of 16 Mbps has been demonstrated with a BER < $3\times10^{-3}$.

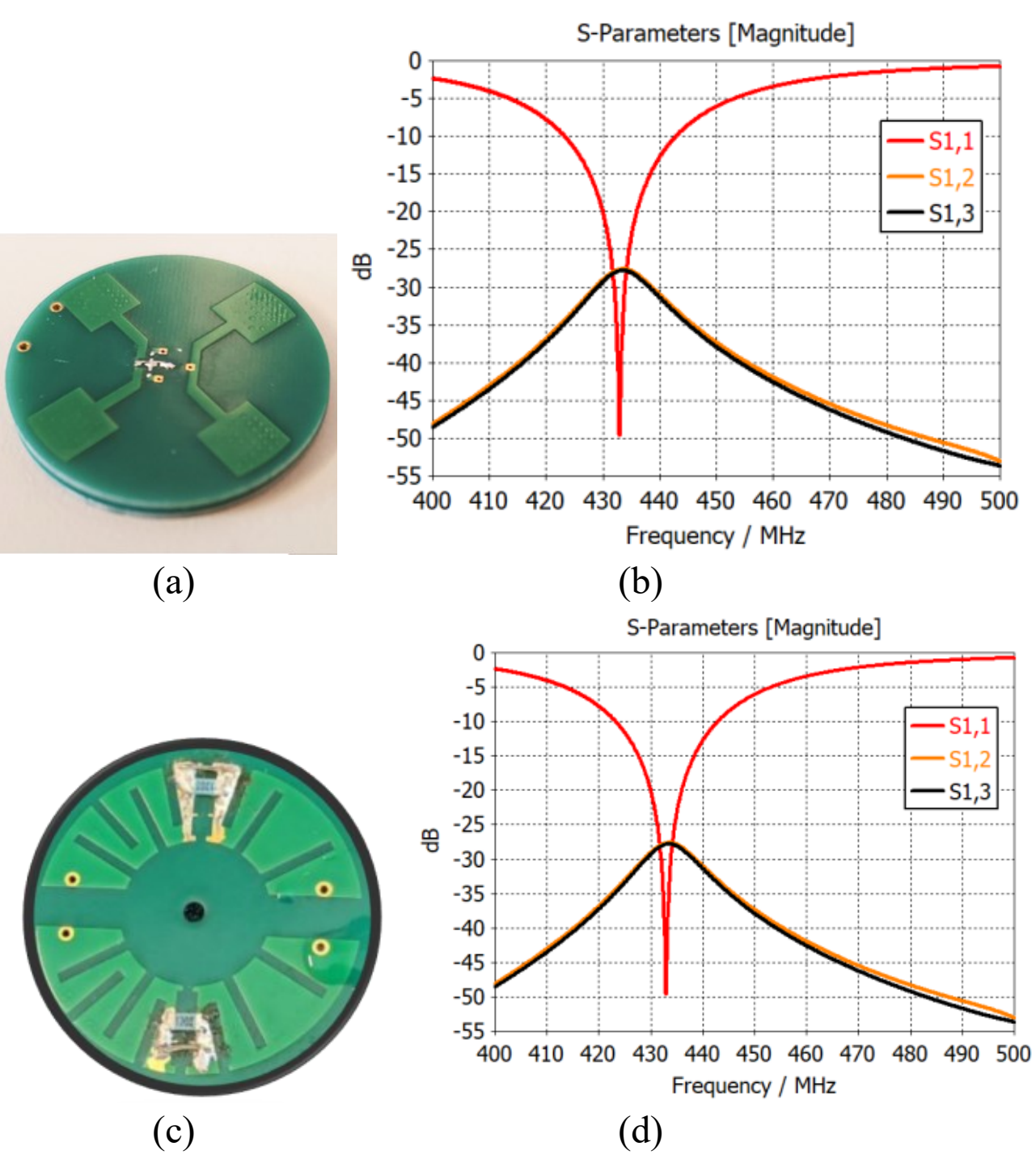

Fig. 3. a) Implant antenna model b) antenna impedance after matching and the coupling to the external reader system antennas c) dual port reader antenna for backscatter d) the reader antenna mutual coupling and the impedance matching with addition of matching unit.

## III. System Demonstration

The implant unit uses NXP NTAG 5 as the NFC two-way communication and also for WPT to support the implant 30 mW power and provide required 2.7 volt voltage line to the ASIC and the implant supporting electronics. The stimulation data is applied via the NFC link to the implant unit, while the NFC is set in continuous charging mode. A microcontroller monitors the line voltage and the implant temperature and sends these telemetry data to the external reader board, specifically the CLEV6630B, acts as the external hub for WPT and bi-directional command/control mechanisms. This board acts as a bridge, translating external commands to the implant and vice versa. All data transmission relies on the NFC link, seamlessly managed by the microcontroller.

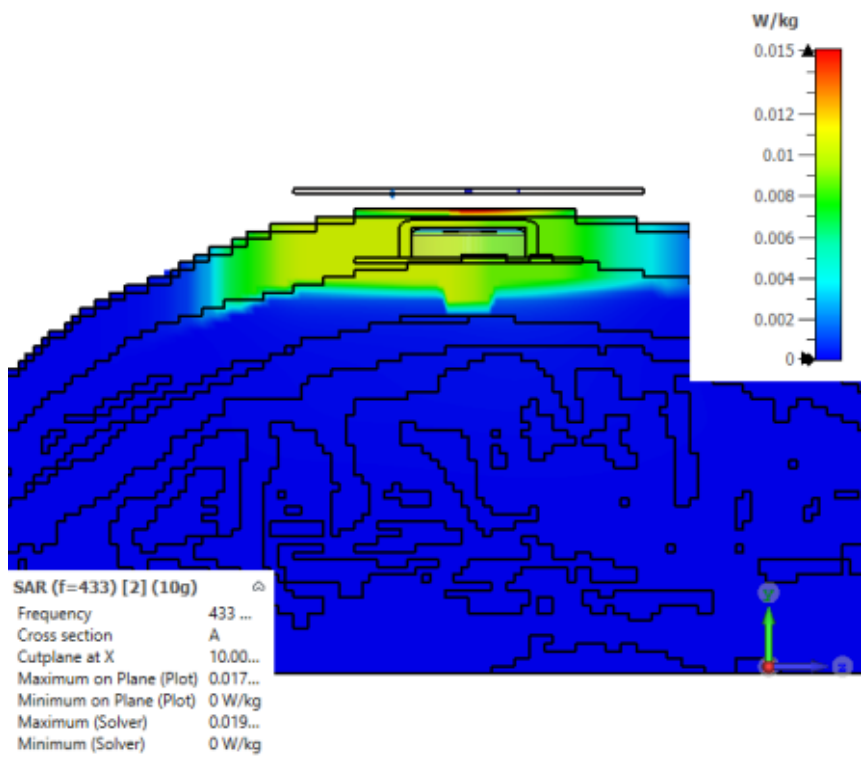

Fig. 4. Computed SAR at 434 MHz using heterogenous human head model, accepted power level 20 mw.

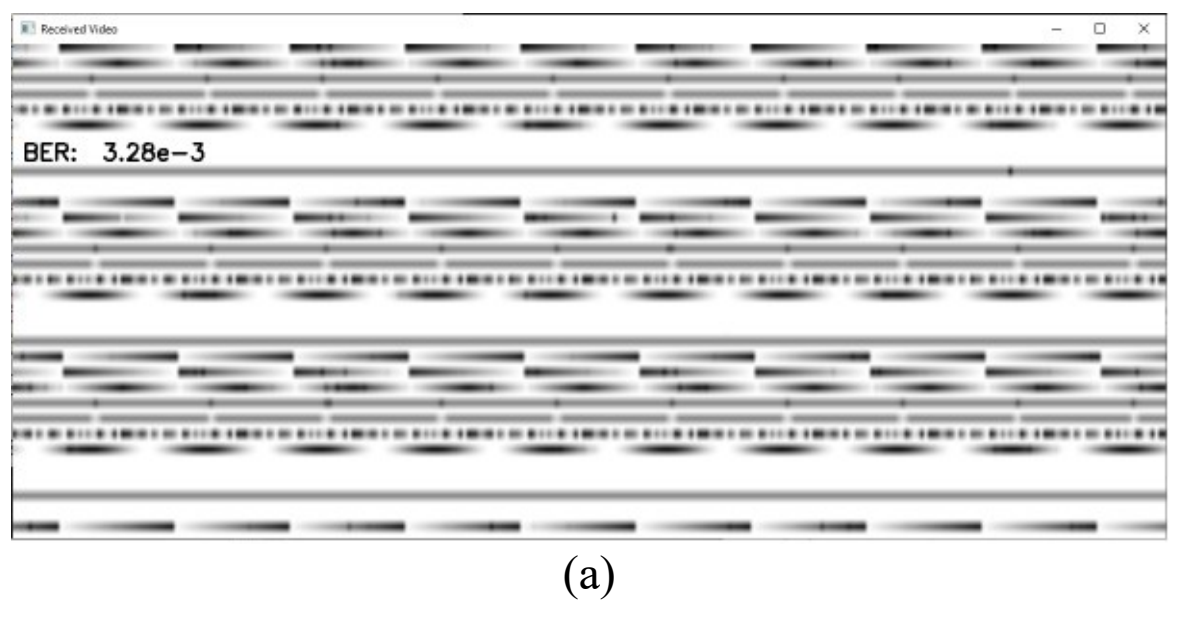

(a)

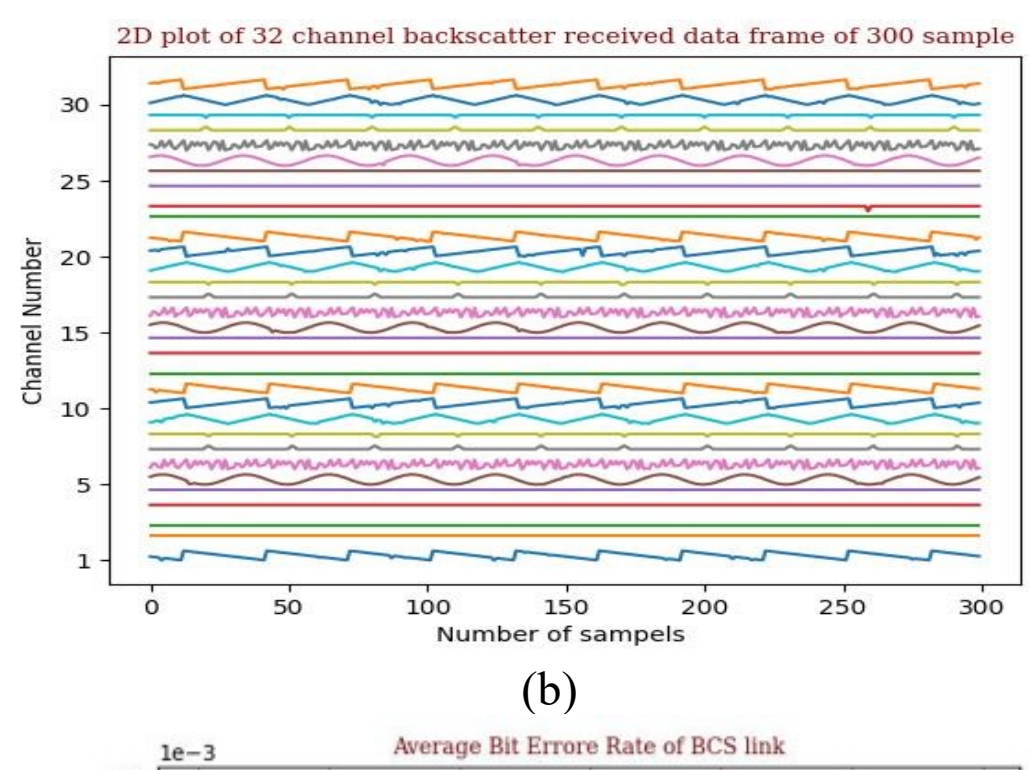

(b)

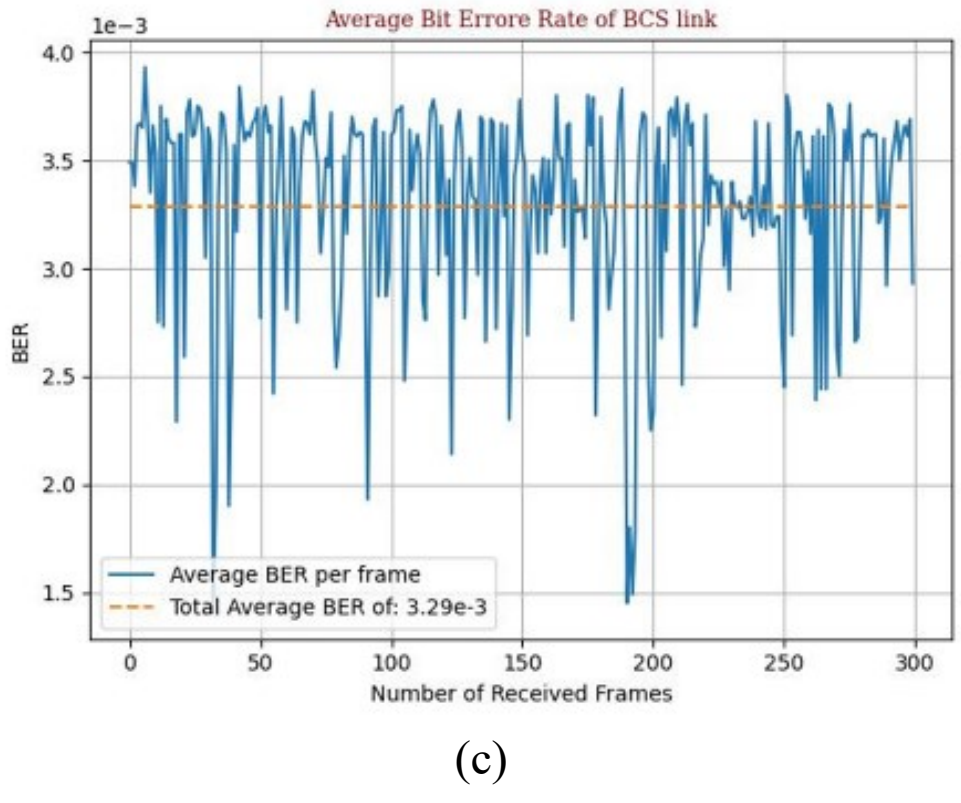

(c)

Fig. 5. a) Received backscattered data for 3 cm separation between implant and external integrated antenna. Each video frame includes 10 period of each channel waveform that any period includs 30 sample of 16 bit, represented as a row in image (32×300×16 bit=153600 bit). In this frame BER=$3.2\times10^{-3}$. b) 2-D plot of received data, each horizontal graph show related channel data. c) calculated BER of each received frame of 153600 bit.

Our integrated system, illustrated in Fig. 5, underwent a loop test. Using an emulator, we generated dummy neural data at 24 Mbps and directed it to the implant module. The external reader antennas are connected to our in-house fabricated reader system that supports backscatter data connectivity of up to 30 Mbps with a standard LAN UDP connection. We've developed a Python code to demonstrate data in real-time. To ensure quality, we've implemented a video representation to gauge transmission accuracy. For technical precision, we based our metrics on 32 distinct emulator channels. Each emits a consistent waveform, facilitating data matching and error pinpointing. As visualized in Fig. 5-b this data is processed and displayed in real-time. Preliminary tests, with the implant 3 cm away from the antenna and a 100 mW external RF transmitter input, shows a promising error rate of only $3.29\times10^{-3}$ average BER.

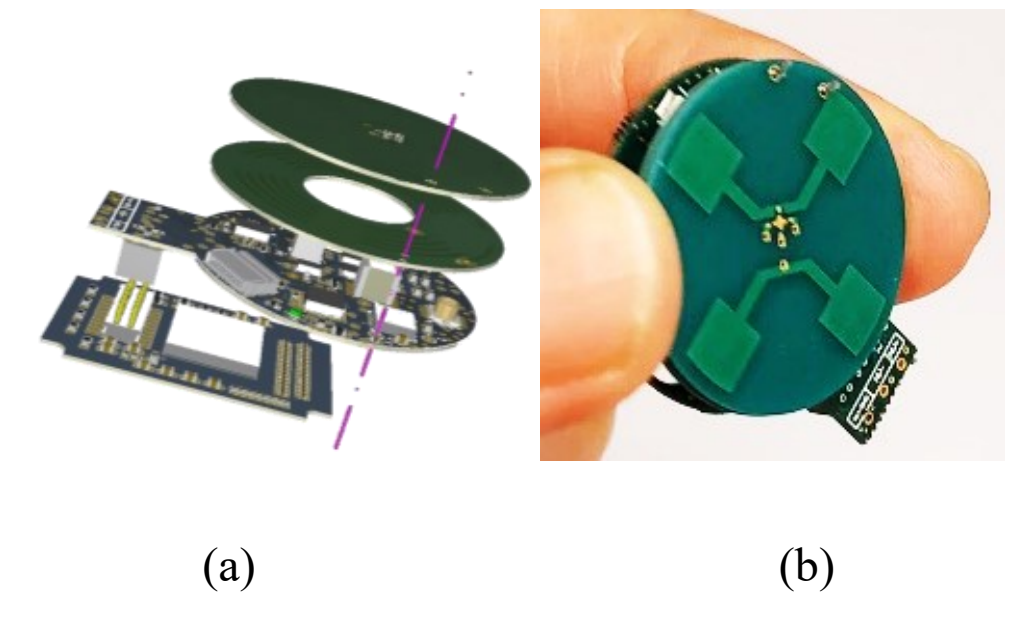

(a) (b)

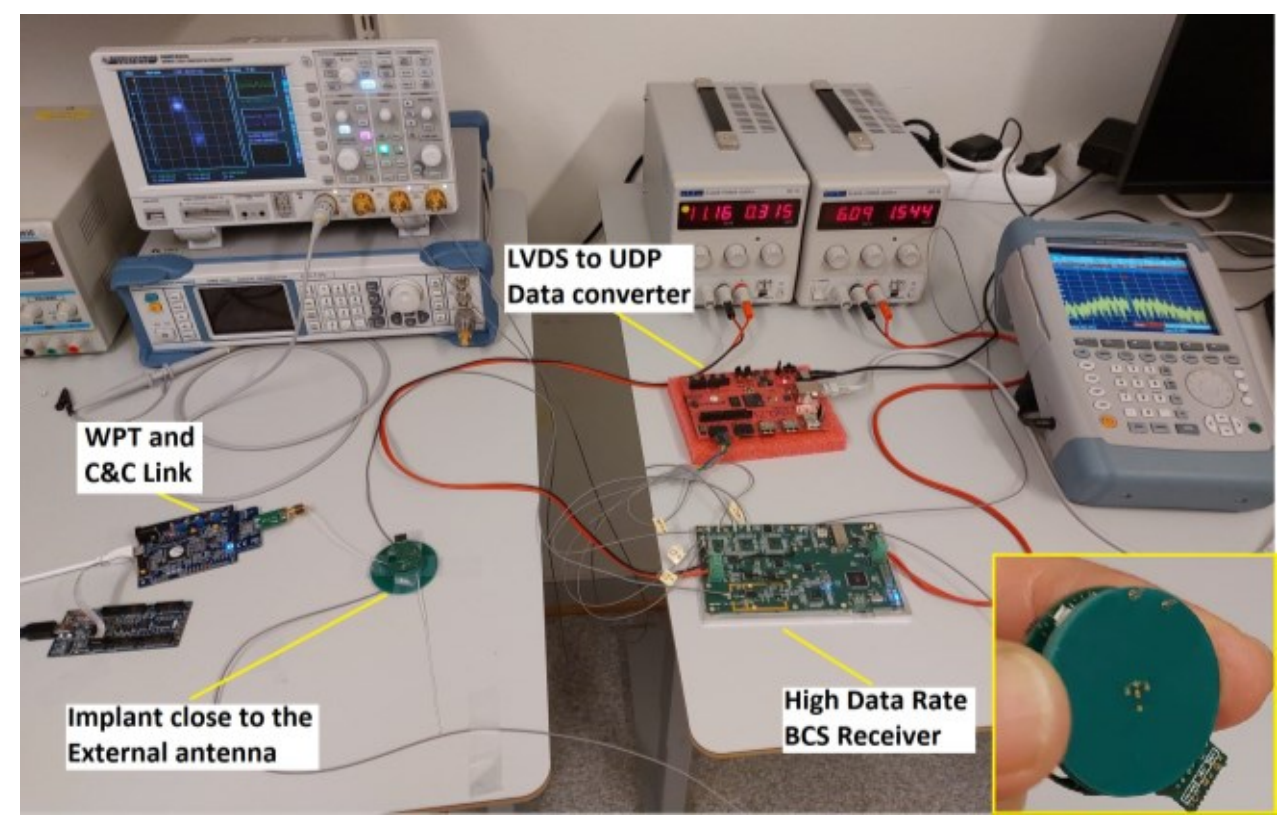

(c)

Fig. 6. a) Implanted part of BMI, (a) integrated manufactured boards, (b) decomposition model and (c) the implanted main board.

## IV. CONCLUSION

In this study, we presented a robust integration of wireless powering via magnetic induction with high-speed data communication tailored for Brain-Machine Interface (BMI) applications. Encased within a compact enclosure, our implant harnesses the standardized frequency of 13.56 MHz, optimized for near-field coupling (NFC) protocols for powering and data telemetry and command, additionally integrated RF backscatter module for 24 Mbps data transmission. Rigorous simulations, spanning both multilayer and heterogeneous head models, verified the safety and efficacy of our approach, consistently maintaining SAR levels below the recommended 2 W/kg threshold.
Our dual-antenna configuration, incorporating bi-static radar mechanisms and unique meander line designs, achieved a remarkable balance between minimized mutual coupling and consistent communication at 434 MHz. System tests, governed by an integrated command/control framework, consistently showed promising results, with an average BER of $3.29\times10^{-3}$. Such achievements, despite complex integration demands, underline the potential of our system in pioneering next-generation wireless BMIs.

## ACKNOWLEDGMENT

The work has been supported by the project Brain-Connected inteRfAce TO machineS (B-CRATOS), (https://www.b-cratos.eu) under grant 965044, Horizon 2020 FET-OPEN.